\documentclass[11pt,a4paper,reqno,twoside]{amsart}

\usepackage{amssymb}
\usepackage{amsthm}
\usepackage{xspace}
\usepackage{amsmath}
\usepackage{setspace}
\usepackage{amscd}
\usepackage{natbib}
\usepackage{amsfonts}
\usepackage{setspace}
\usepackage{textcomp}
\usepackage{enumerate}
\usepackage[english]{babel}
\usepackage{algorithm, algpseudocode}
\usepackage{verbatim} 
\usepackage{soul}
\usepackage{color}
\setul{1ex}{0.8ex}
\definecolor{lightgray}{gray}{0.85}
\sethlcolor{lightgray}

\newtheorem{theorem}{Theorem}[section]
\newtheorem{lemma}[theorem]{Lemma}
\newtheorem{proposition}[theorem]{Proposition}

\newtheorem{definition}[theorem]{Definition}

\algrenewcommand\algorithmicrequire{\textbf{Input:}}
\algrenewcommand\algorithmicensure{\textbf{Output:}}

\begin{document}

\title[Comprehensive Border Bases]
{Comprehensive Border Bases for Zero Dimensional Parametric Polynomial
Ideals}
\author[Abhishek Dubey and  Ambedkar Dukkipati]{Abhishek Dubey and  Ambedkar Dukkipati}
\email{dubey.abhishek@csa.iisc.ernet.in \\ ad@csa.iisc.ernet.in}
\address{Dept. of Computer Science \& Automation\\Indian Institute of Science, Bangalore - 560012}

\begin{abstract}
In this paper, we extend the idea of comprehensive Gr\"{o}bner bases
given by \cite{Weispfenning:1992:ComprehensiveGrobnerbases} to 
border bases for zero dimensional parametric polynomial ideals. For
this, we introduce a notion of comprehensive border 
bases and border system, and prove their existence even in the cases where 
they do not correspond to any term order. We further present algorithms
to compute comprehensive border bases and border system. Finally, 
we study the relation between comprehensive Gr\"{o}bner bases and
comprehensive border bases w.r.t. a term order and give an algorithm 
to compute such comprehensive border bases from comprehensive
Gr\"{o}bner bases. 
\end{abstract}

\maketitle
\section{Introduction}
\label{Introduction}


Since \cite{Weispfenning:1992:ComprehensiveGrobnerbases} introduced and
proved the existence of comprehensive Gr\"{o}bner bases (CGB) for
parametric ideals,
there has been a lot of work done in this direction.
Stability of Gr\"{o}bner bases under specialization was studied in
\citep{Kalkbrener:1997:stabilityofCGB} and based on this,  
an improved algorithm for computing the CGB
was given in \citep{Montes:1999:SpecializationInGrobnerbases}. An
algorithm  for computing CGB from the
Gr\"{o}bner bases of the initial ideal using the computation in the
polynomial ring  over ground field was given in
\citep{SuzukiSato:2006:SimpleAlgoForCGB}.  
Recently an efficient method for computing CGB was given by
\cite{DeepakKapur:2013:EfficientCGBMethod}.  

Border bases, an alternative to Gr\"{o}bner bases, is known to have
more numerical stability as compared to Gr\"{o}bner bases 
\citep{stetter2004numerical}. Recently there have been much interest
in the theory of border bases though they are restricted to zero
dimensional ideals. Characterization of border bases was
given in \citep{KehreinKreuzer:2004:CharacterizationsBB} along with
some parallel results from Gr\"{o}bner bases theory. 
Complexity of border bases detection was studied in
\citep{PrabhanjanAmbedkarDukkipati:2010:NpCompleteBB,AnanthDukkipati:2012:ComplexityOfGrobnerBasisAndBorderBasisDetection}.  
In this paper, we define and study the border system and comprehensive
border bases similar to Gr\"{o}bner system and CGB respectively. 

\subsection*{Contributions}
We define comprehensive border bases (CBB) and border system (BS) on the similar
lines of comprehensive Gr\"{o}bner bases (CGB) and Gr\"{o}bner system (GS).
We show the existence of the comprehensive border bases, even in the
case where they do not correspond to 
any term order. We then provide algorithms for computing border system
and comprehensive border bases.
We show that for a given term order the border form ideal
\citep{KehreinKreuzer:2004:CharacterizationsBB}, 
is same as leading term ideal and, using this fact, we present a
relation between CGB and comprehensive
border bases for a given term order. 

\subsection*{Organization}
The rest of the paper is organized as follows.
In Section~\ref{Preliminaries}, we present preliminaries on
CGB and border bases.
In Section~\ref{Comprehensive Border bases}, we introduce border
system and comprehensive border bases and prove the existence of the
same. 
In Section~\ref{Algo Comprehensive border bases}, we present an
algorithm for computing border system and comprehensive border bases. 
A relation between comprehensive border bases and CGB is presented in
Section~\ref{Realtion between CGB and CBB}.
A brief discussion of CBB over von Neumann regular rings is presented in
Section~\ref{CBB over regular rings}.
Finally a detailed example is given in Section~\ref{Example section} and we
give concluding remarks in Section~\ref{Concluding remarks}.
\section{Background \& Preliminaries}
\label{Preliminaries}
\subsection{Notations}
Throughout this paper, $\Bbbk$ denotes a field, $\bar{\Bbbk}$ denotes
the field extension (algebraic closure) of $\Bbbk$ and $\mathbb{N}$ 
the set of natural numbers. A polynomial ring in indeterminates
$x_{1},\ldots,x_{n}, u_{1},\ldots,u_{m}$ over $\Bbbk$ is denoted by 
$\Bbbk[x_{1},\ldots,x_{n}$, $u_{1},\ldots,u_{m}](=\Bbbk[X,U])$, where
$x_{1},\ldots,x_{n}$ are the main variables and $u_{1},\ldots,u_{m}$
are the parameters. A polynomial ring in the main variables
$x_{1},\ldots,x_{n}$ over $\Bbbk$ is denoted by
$\Bbbk[x_{1},\ldots,x_{n}](=\Bbbk[X])$ and a polynomial ring in the
parameters $u_{1},\ldots,u_{m}$ over $\Bbbk$ is denoted by
$\Bbbk[u_{1},\ldots,u_{m}](=\Bbbk[U])$. 
A monomial
$x_{1}^{\alpha_{1}}\cdots x_{n}^{\alpha_{n}} \in {\mathbb{T}}^n$ is denoted by $x^{\alpha}$,
with the understanding that $x = (x_{1},\ldots,x_{n})$ and
$\alpha = (\alpha_{1},\ldots,\alpha_{n}) \in {\mathbb{N}}^n$.
We denote the set of all monomials in main variables $x_{1},\ldots,x_{n}$
by ${\mathbb{T}}^n$. ${\mathbb{T}}^n_{\ell}$ denotes the set of monomials of degree $\ell$.
When we need to deal with more than one monomials, say $\ell$,
in variables $x_{1},\ldots,x_{n}$, we index these monomials as
$x^{\alpha^{(1)}},x^{\alpha^{(2)}},\ldots,x^{\alpha^{(\ell)}}$.
With respect to a term order $\prec$, we have the leading monomial
($\operatorname{LM}_\prec$), leading coefficient
($\operatorname{LC}_\prec$), leading term ($\operatorname{LT}_\prec$)
and the degree of a polynomial ($\operatorname{deg}_\prec$), 
where $\operatorname{LT}_\prec(f) = \operatorname{LC}_\prec(f)\operatorname{LM}_\prec(f)$ and
$\operatorname{deg}_\prec(f) = \operatorname{deg}_\prec(\operatorname{LM}_\prec(f))$.
If it is clear from the context we will drop the subscript for term order.

A substitution or specialization is a ring homomorphism
\[ \sigma : \Bbbk[x_{1},\ldots,x_{n}, u_{1},\ldots,u_{m}]
\longrightarrow \bar{\Bbbk}[x_{1},\ldots,x_{n}]\] specified by, 
$x_{i} \longmapsto x_{i},\ i=1,\dots,n, u_{j} \longmapsto c_{j},\ j=1,\dots,m,$
where $c_{1},\ldots,c_{m} \in \bar{\Bbbk}$.
By this the specialization $\sigma$ can be uniquely identified by
$\sigma = (c_{1},\ldots,c_{m}) \in \bar{\Bbbk}^m$. 
A set of specializations is termed as a condition in parameters
$u_{1},\ldots,u_{m}$ and is denoted by $\gamma$.
For an ideal $\mathfrak{a} \subseteq \Bbbk[U]$,
$\mathcal{V}(\mathfrak{a}) \subseteq \bar{\Bbbk}^m$ denotes the
algebraic set of $\mathfrak{a}$, i.e., 
\begin{displaymath}
\mathcal{V}(\mathfrak{a}) = \{(c_{1},\ldots,c_{m}) \in \bar{\Bbbk}^m
\mid  f \in \mathfrak{a}, f(c_{1},\ldots,c_{m}) = 0\}.
\end{displaymath}

\subsection{Comprehensive Gr\"{o}bner bases}
\begin{definition}
Let $F$ be a finite set of generators of an ideal $\mathfrak{a} \subseteq \Bbbk[U][X]$ and $S$ be a subset of $\bar{\Bbbk}^m$. A finite
subset $G$ in $\Bbbk[U][X]$ is called a comprehensive Gr\"{o}bner basis on $S$ for $F$, if $\sigma(G)$ is a Gr\"{o}bner basis for the ideal
$\sigma(F)$ in $\bar{\Bbbk}[X]$ for any specialization $\sigma \in S$. If $S = \bar{\Bbbk}^m$, then $G$ is called a
comprehensive Gr\"{o}bner basis (CGB) for $F$.
\end{definition}

%
Let $G_{= 0}$ = $\{g_{1}, \ldots, g_{s}\}$ and
$G_{\neq 0}$ = $\{g'_{1}, \ldots, g'_{t}\}$ be a finite set of polynomial equations and polynomial inequalities
in $\Bbbk[U]$ respectively.
Then a condition, $\gamma \subseteq \bar{\Bbbk}^m$, can also be represented as $\mathcal{V}(G_{= 0}) \setminus \mathcal{V}(G_{\neq 0})$.

\begin{definition}
Let $F$ be a subset of $\Bbbk[U][X]$, $\gamma_{1},\ldots, \gamma_{t}$ be
conditions in $\bar{\Bbbk}^m$, 
$G_{1},\ldots, G_{t}$ be subsets of $\Bbbk[U][X]$, and $S$ be a subset
of $\bar{\Bbbk}^m$ such that 
$S \subseteq \gamma_{1} \cup \ldots \cup \gamma_{t}$. A finite set $\mathcal{G}$ =
\{$(G_{1}, \gamma_{1}),\ldots, (G_{t}, \gamma_{t})$\} is called a Gr\"{o}bner
system on $S$ for $F$, if $\sigma(G_{i})$ is a Gr\"{o}bner basis for the 
ideal $\sigma(F)$ in $\bar{\Bbbk}[X]$ for any specialization $\sigma
\in \gamma_{i}$, $i = 1,\ldots, t$. If $S = \bar{\Bbbk}^m$, then
$\mathcal{G}$ is called a Gr\"{o}bner system $(GS)$ for $F$. 
\end{definition}

For any distinct pairs $(G_{i}, \gamma_{i}),(G_{j}, \gamma_{j}) \in \mathcal{G}$ such that $\gamma_{i} \cap \gamma_{j} = \gamma_{\ell}$,
we can replace $(G_{i}, \gamma_{i}),(G_{j}, \gamma_{j}) \in \mathcal{G}$ with
$(G_{i}, \gamma_{i} \setminus \gamma_{\ell}),(G_{j}, \gamma_{j} \setminus \gamma_{\ell}), (G_{\ell}, \gamma_{\ell})$, where
$G_{\ell}$ is either $G_{i}\ \mathrm{or}\ G_{j}$.
So without any loss of generality we will assume that the conditions,
$\gamma_{1},\ldots, \gamma_{t}$, are pairwise disjoint.

A CGB $G$ for $F$ is called faithful, if
$G \subseteq \langle F\rangle$. A Gr\"{o}bner system
$\mathcal{G} = \{(G_{1}, \gamma_{1}),\ldots, (G_{t}, \gamma_{t})\}$ for $F$ is
called faithful if every element of $G_{i}$, is also in $\langle
F\rangle$, for $i = 1,\ldots, t$. 

The first algorithm to compute CGB was given by
\citet{Weispfenning:1992:ComprehensiveGrobnerbases}. The central idea
in this approach is to generate a finite partition (case distinction) of
$\operatorname{spec}(\Bbbk[U])$ such that, inside each set of the partition the leading term
ideal remains same for all specialization satisfying the condition of
the set. The polynomials are `conditionally colored' to
calculate the conditional Gr\"{o}bner bases under a condition.
Under a given partition 
(condition) each term is colored according to the value of its
coefficient as follows: the terms with coefficients that vanish 
under the condition are colored as `green' and terms with coefficients
that do not vanish under the condition are colored `red'. When the
condition is not sufficient to color the term as red or green, we
color the term as `white'. A polynomial with all terms colored is
called a colored polynomial. The Gr\"{o}bner bases computations are then done with
respect to conditional leading term of a colored polynomial, to make
sure we generate a conditional Gr\"{o}bner bases that is not outside
the original ideal. The conditions are further refined in case the
Gr\"{o}bner bases computation generate a white leading term polynomial.
Gr\"{o}bner bases along with each partition is called
Gr\"{o}bner system and the union of all the Gr\"{o}bner bases of Gr\"{o}bner
system gives CGB. For a detailed
exposition one can refer to
\citep{WillMatthisDunn:1995:CGBThesis}.
\subsection{Border bases in $\Bbbk[x_{1},\ldots,x_{n}]$}
\begin{definition} A finite set of terms $\mathcal{O} \subset {\mathbb{T}}^n$ is called an order ideal if it is
closed under forming divisors i.e. for $x^{\alpha} \in {\mathbb{T}}^n$ if, $x^{\beta} \in \mathcal{O}$ and $x^{\alpha} | x^{\beta}$ then
it implies $x^{\alpha} \in \mathcal{O}$.
\end{definition}

\begin{definition} Let $\mathcal{O} \subset {\mathbb{T}}^n$ be an order ideal. The border of $\mathcal{O}$ is the set
$\partial \mathcal{O} = ({\mathbb{T}}_{1}^n\mathcal{O}) \setminus \mathcal{O}$ =
$(x_{1}\mathcal{O} \cup \ldots \cup x_{n}\mathcal{O}) \setminus \mathcal{O}$.
The first border closure of $\mathcal{O}$ is defined as the set $\mathcal{O} \cup \partial \mathcal{O}$ and it is denoted by
$\overline{\partial \mathcal{O}}$.\\
Note that $\overline{\partial \mathcal{O}}$ is also an order ideal.
\end{definition}
By convention for $\mathcal{O} = \emptyset$ we set $\partial
\mathcal{O} = 1$. 

\begin{definition} Let $\mathcal{O} = \{x^{\alpha^{(1)}}, \ldots, x^{\alpha^{(s)}}\} \subset {\mathbb{T}}^n$ be an order ideal, and
let $\partial \mathcal{O} = \{x^{\beta^{(1)}}, \ldots, x^{\beta^{(t)}}\}$ be it’s border. A set of polynomials B =
$\{b_{1}, \ldots, b_{t}\} \subset \Bbbk[X]$ is called an $\mathcal{O}$-border prebasis if the polynomials have the form
$b_{j} = x^{\beta^{(j)}} - \sum_{i=1}^{s} c_{ij}x^{\alpha^{(i)}}$, where $c_{ij} \in \Bbbk\ \mathrm{for}\ 1 \leq i \leq s$ and
$1 \leq j \leq t$.
\end{definition}

\begin{definition} Let $\mathcal{O} = \{x^{\alpha^{(1)}}, \ldots, x^{\alpha^{(s)}}\} \subset {\mathbb{T}}^n$ be an order ideal and
B = $\{b_{1}, \ldots, b_{t}\}$ be an $\mathcal{O}$-border prebasis consisting of polynomials in $\mathfrak{a} \subseteq \Bbbk[X]$.
We say that the set B is an $\mathcal{O}$-border bases of $\mathfrak{a}$ if the residue classes of
$x^{\alpha^{(1)}}, \ldots, x^{\alpha^{(s)}}$ form a $\Bbbk$-vector space basis of $\Bbbk[x_{1},\ldots, x_{n}]/\mathfrak{a}$.
\end{definition}

Now we give a brief account of characterization of border bases given in \citep{KehreinKreuzer:2004:CharacterizationsBB}.

Index of $x^{\beta} \in {\mathbb{T}}^n$ is defined as
$ind_{\mathcal{O}}(x^{\beta}) = \operatorname{min}\{r \geqslant 0 | x^{\beta} \in \overline{\partial^{r} \mathcal{O}} \}$.
Given a non-zero polynomial $f = c_{1}x^{\alpha^{(1)}}+\ldots+c_{s}x^{\alpha^{(s)}} \in \mathfrak{a} \subseteq \Bbbk[X]$,
where $c_{1},\ldots, c_{s} \in \Bbbk \setminus \{0\}$ and $x^{\alpha^{(1)}},\ldots, x^{\alpha^{(s)}} \in {\mathbb{T}}^n$,
we order the terms in the support of $f$ such that
$ind_{\mathcal{O}}(x^{\alpha^{(1)}}) \geq ind_{\mathcal{O}}(x^{\alpha^{(2)}}) \geq \ldots \geq ind_{\mathcal{O}}(x^{\alpha^{(s)}})$.
Then we call $ind_{\mathcal{O}}(f) = ind_{\mathcal{O}}(x^{\alpha^{(1)}})$ the index of $f$.

\begin{definition}
 Given a polynomial f $\in \mathfrak{a} \subseteq \Bbbk[X]$, there is a representation of
 $f = c_{1}x^{\alpha^{(1)}}+\ldots+c_{s}x^{\alpha^{(s)}}\ \mathrm{with}\ c_{1},\ldots, c_{s} \in \Bbbk \setminus\{0\}$ and
 $x^{\alpha^{(1)}},\ldots, x^{\alpha^{(s)}} \in {\mathbb{T}}^n$ such that
$ind_{\mathcal{O}}(x^{\alpha^{(1)}}) \geq ind_{\mathcal{O}}(x^{\alpha^{(2)}}) \geq \ldots \geq ind_{\mathcal{O}}(x^{\alpha^{(s)}})$.
    \begin{enumerate}
    \item The polynomial $BF_{\mathcal{O}}(f) = \sum\limits_{\substack{i = 1 \\ ind(x^{\alpha^{(i)}}) = ind(f )}}^s c_{i}x^{\alpha^{(i)}}$ is called the
    border form of $f$ with respect to $\mathcal{O}$. For $f = 0$, we let $BF_{\mathcal{O}}(f) = 0$.
    \item Given an ideal $\mathfrak{a}$, the ideal $BF_{\mathcal{O}}(\mathfrak{a}) = \{BF_{\mathcal{O}}(f) \mid f \in \mathfrak{a}$\}
    is called the border form ideal of $\mathfrak{a}$ with respect to $\mathcal{O}$.
    \end{enumerate}
\end{definition}

\begin{definition}
$F$ is $\mathcal{O}$-border bases of $\mathfrak{a} \subseteq \Bbbk[X]$ if and only if one of the following equivalent conditions is satisfied:
    \begin{enumerate}
    \item For all $f \in \mathfrak{a}$, $\operatorname{supp}(BF_{\mathcal{O}}(f))$ is contained in ${\mathbb{T}}^n \setminus \mathcal{O}$.
    \item $BF_{\mathcal{O}}(\mathfrak{a}) = (x^{\beta^{(1)}}, \ldots, x^{\beta^{(t)}})$, where
    $x^{\beta^{(i)}} \in \{ BF_{\mathcal{O}}(f_{i}) | f_{i} \in F \}$.
    \end{enumerate} 
\end{definition}

Reduced Gr\"{o}bner bases is unique for an ideal with respect to a given
term order. While the monomial ordering in the Gr\"{o}bner bases theory is given by term ordering, in the border bases theory it can be
given by the shortest distance from the order ideal.

\noindent
\begin{proposition}
For a given term order and an ideal there exists a unique reduced Gr\"{o}bner bases and a unique border bases.
\end{proposition}
The proof of above proposition follows from the following results. 
\begin{theorem}[Macaulay-Buchberger Basis Theorem \citep{Buchberger:1965:thesis}]
Let $G = \{g_1, \ldots, g_t\}$ be a Gr\"{o}bner basis for an ideal $\mathfrak{a} \subseteq \Bbbk[x_1,\ldots,x_n]$. 
A basis for the vector space $\Bbbk[x_1,\ldots,x_n]/\langle \mathfrak{a} \rangle$ is given by
$S = \{ x ^\alpha + \mathfrak{a} : \mathrm{LM}(g_i)\nmid x^\alpha, i=1,\ldots,t\}$.
\end{theorem}

\begin{theorem}\label{Unique Border and reduced Grobner bases}[\citep{KehreinKreuzerRobbiano:2005:BBSurvey}]
Let $\mathcal{O} = \{x^{\alpha^{(1)}}, \ldots, x^{\alpha^{(s)}}\} \subset {\mathbb{T}}^n$ be an order ideal, let $\mathfrak{a}$ be a
zero-dimensional ideal in $\Bbbk[X]$, and assume that the residue classes of the elements in $\mathcal{O}$ form a $\Bbbk$-vector
space basis of $\Bbbk[X]/\mathfrak{a}$:
\begin{enumerate}
 \item There exists a unique $\mathcal{O}$-border basis $G$ of $\mathfrak{a}$.
 \item Let $G'$ be an $\mathcal{O}$-border prebasis whose elements are in $\mathfrak{a}$. Then $G'$ is the $\mathcal{O}$-border basis of
 $\mathfrak{a}$.
 \item Let $\Bbbk$ be the field of definition of $\mathfrak{a}$. Then we have $G \subset \Bbbk[X]$.
\end{enumerate}
\end{theorem}
\endproof

As we have mentioned earlier, given an ideal there exists border bases
that  do not correspond to Gr\"{o}bner bases for any term ordering. 
An example of such a border basis is given in \citep{KehreinKreuzer:2006:ComputingBB}.


\section{Comprehensive Border bases}
\label{Comprehensive Border bases}
First we define the notion of `scalar' border bases.
\begin{definition}\label{SBB}
 Let $\mathfrak{a} \subseteq \Bbbk[X]$ be an ideal, for $\{b_{1},
 \ldots,b_{t}\} \subset \mathfrak{a}$, 
 $\mathcal{O} \subset \mathbb{T}^n$ an order ideal and
 $c_{1},\ldots,c_{t} \in \Bbbk$, $\{c_{1}b_{1}, \ldots,c_{t}b_{t}\}$
 is called scalar $\mathcal{O}$-border bases of $\mathfrak{a}$ if
 $\{b_{1}, \ldots,b_{t}\}$ is an $\mathcal{O}$-border basis of
 $\mathfrak{a}$.  
\end{definition}
We need the notion of scalar border bases because our aim is to
construct border bases for the ideals in $\bar{\Bbbk}[X]$ that result
from specialization of the ideals in $\Bbbk[U][X]$. 
As the coefficient space changes from the ring $\Bbbk[U]$ to the field $\bar{\Bbbk}$
 under a specialization, we will not be able to get the monic border form terms for the border bases in $\bar{\Bbbk}[X]$ from the
 polynomials in $\Bbbk[U][X]$.
Now we define the notion of border system and comprehensive border
bases. 
\begin{definition}\label{BS}
Consider a zero dimensional ideal $\mathfrak{a} \subseteq \Bbbk[X,U]$ and
$S \subseteq \bar{\Bbbk}^m$. Let $\gamma_{\ell}$ is a
condition, $\mathcal{O}_{\ell}$ is a order ideal and $B_{\ell} \subset
\Bbbk[X,U]$, for $\ell =1,\ldots,L$. Then the set  
$\mathcal{B} = \{(\gamma_{\ell}, \mathcal{O}_{\ell}, B_{\ell}) : \ell =
1,\ldots,L \}$ is said to be a border system on $S$ for $\mathfrak{a}$
if, 
\begin{enumerate} 
  \item $\gamma_{i} \cap \gamma_{j} = \emptyset$ for all $i, j$ such that $1\leq i,j\leq L,\ i \neq j$,
  \item $S \subseteq \bigcup\limits_{i = 1}^{L}\gamma_{i}$, and
  \item for every specialization $\sigma \in \gamma_{\ell}$, $\sigma(B_{\ell})$ is a scalar $\mathcal{O}$-border bases of $\sigma(\mathfrak{a})$
	in $\bar{\Bbbk}[X]$.
 \end{enumerate}
If $S = \bar{\Bbbk}^m$, then $\mathcal{B}$ is called border system for
$\mathfrak{a}$. 
\end{definition}

\begin{definition}\label{CBB}
  A finite set $B \subset \Bbbk[X,U]$ is said to be a comprehensive border basis on
$S \subseteq \bar{\Bbbk}^m$ for a zero dimensional ideal $\mathfrak{a} = \langle B\rangle$ if for all specializations
$\sigma \in S$, there exists an order ideal $\mathcal{O} \subset \mathbb{T}^n$ such that, $\sigma(B)$ is a
scalar $\mathcal{O}$-border basis of $\sigma(\mathfrak{a}) \subseteq \bar{\Bbbk}[X]$.
If $S = \bar{\Bbbk}^m$, then $B$ is called comprehensive border basis
for $\mathfrak{a}$. 
\end{definition}

A  border system $\mathcal{B} = \{(\gamma_{\ell}, \mathcal{O}_{\ell},
B_{\ell}) : \ell = 1,\ldots,L \}$ for an ideal $\mathfrak{a}$ is called faithful, 
if in addition, every element of $B_{\ell},\: \ell = 1,\ldots,L$ is also in $\mathfrak{a}$.
A comprehensive border basis ${B}$ for $\mathfrak{a}$ is called faithful, if in addition, every element of $B$ is also in $\mathfrak{a}$.

\begin{definition}\label{ConditionalBB} Let $\sigma$ be a specialization and $\mathfrak{a} \subseteq \Bbbk[X,U]$ be an ideal,
then $B \subset \mathfrak{a}$ is
called a conditional border basis of $\mathfrak{a}$ under $\sigma$ if,
$\sigma(B)$ is an $\mathcal{O}$-border basis of
$\sigma(\mathfrak{a}) \subseteq \bar{\Bbbk}[X]$ for an order ideal
$\mathcal{O} \subset \mathbb{T}^n$. In particular, we say, $B$ is a
$\sigma$-conditional $\mathcal{O}$-border basis for
$\sigma(\mathfrak{a}) \subseteq \bar{\Bbbk}[X]$.  
\end{definition}

Let $\mathfrak{a} \subseteq \Bbbk[x_{1},\ldots,x_{n}$, $u_{1},\ldots,u_{m}]$ be an ideal and
$\mathfrak{a}' = \mathfrak{a} \cap \Bbbk[u_{1},\ldots,u_{m}]$. For
each specialization $\sigma \notin
\mathcal{V}(\mathfrak{a}')$, we have $\sigma(\mathfrak{a}) = \langle 1 \rangle
= \Bbbk[x_{1},\ldots,x_{n}]$ \citep{SuzukiSato:2006:SimpleAlgoForCGB}.
We use this fact along with Lemma~\ref{FiniteVarietyClaim} for zero dimensional
ideals to show the existence of comprehensive border bases. 

\begin{lemma}\label{FiniteVarietyClaim}
 If $\mathfrak{a} \subseteq \Bbbk[x_{1},\ldots,x_{n}$, $u_{1},\ldots, u_{m}]$ is zero-dimensional ideal, then
 $\mathfrak{a}' = \mathfrak{a} \cap \Bbbk[u_{1},\ldots,u_{m}]$ is also a zero dimensional ideal.
\end{lemma}
\proof
Let $G$ be a Gr\"{o}bner basis w.r.t. an elimination (term) order $\le$ with $\{x_{1},\ldots, x_{n}\} \le \{u_{1},\ldots,u_{m}\}$ and
a lex term order within the main variables and parameters.
If $\mathfrak{a}$ is a zero-dimensional ideal then for each $x_{i} (\mathrm{or}\ u_{j})$, $\exists g_{i} \in G$ such that
$\operatorname{LT}(g_{i}) = x_{i}^{\ell}$ (or $u_{i}^{\ell}$ respectively), for some $\ell \in \mathbb{N}$~\citep{Adams:1994:introtogrobnerbasis}.
Also for each $u_{i}\ \exists g_{i} \in G$ such that
$\operatorname{LT}(g_{i}) = u_{i}^{j}$ and $\operatorname{supp}(g_{i}) 
\subset \Bbbk[u_{1},\ldots,u_{m}]$. The Gr\"{o}bner bases for $\mathfrak{a}'$ is $G \cap \Bbbk[u_{1},\ldots, u_{m}]$ \citep{Adams:1994:introtogrobnerbasis}
and we denote it by $G'$.
Now for each $u_{i}\ \exists g_{i} \in G'$ such that $\operatorname{LT}(g_{i}) = u_{i}^{j}$ implying $\mathfrak{a}'$ is a zero dimensional ideal.
\endproof

It is easy to see that if $\mathfrak{a} \subseteq \Bbbk[X,U]$ is a zero
dimensional ideal, then $\sigma(\mathfrak{a}) \subset \Bbbk[X]$ is
also a zero dimensional ideal. Hence there exists a border basis for
$\sigma(\mathfrak{a})$. 

\begin{proposition}\label{CBBExistenceProof}
 For a zero dimensional ideal $\mathfrak{a} \subseteq \Bbbk[X,U]$,
 border system and hence comprehensive border basis always exists. 
\end{proposition}
\proof
Let $F$ be the given set of generators of the ideal $\mathfrak{a}$
in $\Bbbk[X,U]$ and $\sigma \in \bar{\Bbbk}^m$ be a
specialization. Let $\mathfrak{a}' = \mathfrak{a} \cap
\Bbbk[u_{1},\ldots,u_{m}]$ then by Lemma~\ref{FiniteVarietyClaim} we
have $\mathcal{V}(\mathfrak{a}')$ is finite. Now, either $\sigma \in
\mathcal{V}(\mathfrak{a}')$ or $\sigma 
\notin \mathcal{V}(\mathfrak{a}')$. In the first case, for each
$\sigma \in \mathcal{V}(\mathfrak{a}')$ we can compute conditional
border basis for the ideal generated by $F$ as we can exactly
determine which coefficients vanishes and which does not under any
specialization $\sigma$. In the second case, where $\sigma \notin
\mathcal{V}(\mathfrak{a}')$, for all $\sigma \in \bar{\Bbbk}^m
\setminus \mathcal{V}(\mathfrak{a}')$, $F$ can be considered as a
conditional border basis as $\sigma(F) = \sigma(\mathfrak{a}) = \langle 1\rangle$.

There exists a border system $\mathcal{B}$ that is constructed as,
for each $\sigma_{i} \in \mathcal{V}(\mathfrak{a}')$ with $B_{i}$ 
calculated as conditional $\mathcal{O}_{i}$-border basis we add
$(\sigma_{i}, \mathcal{O}_{i}, B_{i})$ to $\mathcal{B}$. For all $\sigma \in \bar{\Bbbk} \setminus \mathcal{V}(\mathfrak{a}')$
we add the tuple $(\sigma, \emptyset, F')$, where $F' = F$ (or $F'$ can also be the finite set of generators of the ideal $\mathfrak{a}'$).

Consider $\varPhi$ as a function which maps an element in $\bar{\Bbbk}^m$ to the corresponding order ideal in the border system
computed above.
Now CBB can be computed as the union of all the conditional border bases from the border system along with $\varPhi$.
\endproof



In Section~\ref{Realtion between CGB and CBB}, we give a method to
construct CBB from CGB, by which one can establish the existence of
CBB. But the significance of the Proposition~\ref{CBBExistenceProof}
arises from the fact that it establishes the existence of border
system and CBB that need not correspond to any term order. Before
we study the relation between CGB and CBB we present the algorithms to
construct border system and CBB.  
\section{Algorithm}
\label{Algo Comprehensive border bases}
Here we briefly recall the `coloring terminology' used by
\cite{Weispfenning:1992:ComprehensiveGrobnerbases}.
Let t = $ax^{\alpha} \in \Bbbk(U)[X]$ be a term, where $\Bbbk(U)$ is the field of fractions of $\Bbbk[U]$, $a \in \Bbbk(U)$ and
$x^{\alpha} \subset \mathbb{T}^n$ is a monomial in main variables $x_{1},\ldots,x_{n}$.
Then color of t is green if $\sigma(a)$ is zero, otherwise it is red.
The basic idea for coloring as given by
\cite{Weispfenning:1992:ComprehensiveGrobnerbases} was to do the
computations for an ideal $\sigma(\mathfrak{a}) \subset \Bbbk[X]$ and
still be able to generate and deal with the polynomials in
$\mathfrak{a} \subseteq \Bbbk(U)[X]$. As we know the exact
specialization for which we need to compute border bases, we will be
able to color the every term of the polynomials in $\mathfrak{a}
\subset \Bbbk(U)[X]$ as either red or green. Now we can use any border
bases algorithm and compute the conditional border bases for
$\sigma(\mathfrak{a}) \subset \Bbbk[X]$ with respect to the red (non
zero) term of the colored polynomials making sure that the modified
polynomials belongs to $\mathfrak{a}$. 

The proof of correctness of the colored border bases algorithm is implied by the proof
of correctness for the main border bases algorithm. As the proof holds for the
red terms of the polynomials and the green terms vanishes under substitution.



We list a procedure to compute border system in Algorithm~\ref{Border System Algorithm}.
\begin{algorithm}[!ht]
 \caption{Border System}\label{Border System Algorithm}
    \begin{algorithmic}[1]
    \Require $F \subset \Bbbk[X,U]$, such that $\langle F\rangle$ is a zero dimensional ideal.
    \Ensure A border system $\mathcal{B}$ for $F$.
     \State Compute $\mathcal{V}(\mathfrak{a}') \in \bar{\Bbbk}^{m}$, where $\mathfrak{a}' = \langle F\rangle \cap \Bbbk[U]$.
     \State $F' = F \setminus \Bbbk[U] \subset \Bbbk(U)[X]$.
	\For{each $\sigma \in \mathcal{V}(\mathfrak{a}')$}	    
	    \State \parbox[t]{\dimexpr\linewidth-\algorithmicindent}{Color $F'$ with $\sigma$, remove the polynomials with all
		    green terms, call $F'_{\sigma}$.}
	    \State\label{coloredBBStep} \parbox[t]{\dimexpr\linewidth-\algorithmicindent}{Compute conditional $\mathcal{O}$-border bases $B$ for
		    $\langle F'_{\sigma}\rangle$ by the colored version of the border bases algorithm in $\Bbbk(U)[X]$. \strut}
	    \State\label{scalarBBstep}\parbox[t]{\dimexpr\linewidth-\algorithmicindent}{Convert the conditional $\mathcal{O}$-border bases $B$
		    generated above to conditional scalar $\mathcal{O}$-border bases $B'$ in $\Bbbk[U][X]$}
	    \State Update $\mathcal{B} = \mathcal{B} \cup \{(\sigma,\mathcal{O},B')\}$
	\EndFor
    \State\label{outsideVarietyStep} Update $\mathcal{B} = \mathcal{B} \cup \{(\{\bar{\Bbbk}^{m} \setminus \mathcal{V}(\mathfrak{a}')\}, \emptyset, F)\}$
    \State\label{compressBSstep} Compress $\mathcal{B}$
    \State Return $\mathcal{B}$
 \end{algorithmic}
\end{algorithm}

The termination of the algorithm is obvious by
Lemma~\ref{FiniteVarietyClaim}.
Note that Step~\ref{compressBSstep} 
combines two elements of the border system $\mathcal{B}$, when 
they differ only in the specialization of the parameters.
This optional
step is shown in the example we present in
Section~\ref{Example section}. The correctness of the border system algorithm
follows from Proposition~\ref{CBBExistenceProof}. 

We must note that the nature of border system generated by Algorithm~\ref{Border System Algorithm} depends on the nature of the
border bases algorithm used in Step~\ref{coloredBBStep}. Using a border bases (colored version) algorithm we can generate
border system and hence the CBB that do not correspond to any term order.


We need the following definition of vanishing polynomial with respect to
a specialization $\sigma$ and a finite condition, $\gamma$.
\begin{definition}
Let $\sigma = (c_{1}, \ldots, c_{m}) \in \bar{\Bbbk}^{m}$ be a
specialization of parameters 
$(u_{1}, \ldots, u_{m})$, then the vanishing polynomial at $\sigma$ is
defined as
\begin{displaymath}
 f_{\sigma} = (u_{1} - c_{1})^2+ \ldots + (u_{m} - c_{m})^2.
\end{displaymath} 
For a finite condition $\gamma = \{\sigma_{1}, \sigma_{2}, \ldots,
\sigma_{\ell}\}$ we define vanishing polynomial w.r.t. $\gamma$ as,
\begin{displaymath}
 f_{\gamma} = f_{\sigma_{1}}f_{\sigma_{2}}\ldots f_{\sigma_{\ell}}.
 \end{displaymath}
\end{definition}

We list a method to compute CBB in Algorithm~\ref{Comprehensive border bases Algorithm}.

\begin{algorithm}[H]
 \caption{Comprehensive border bases}\label{Comprehensive border bases Algorithm}
    \begin{algorithmic}[1]
    \Require The border system $\mathcal{B} = \{(\gamma_{\ell},\mathcal{O}_{\ell},B_{\ell}) ;\ \ell = 1,\ldots,L \}$ of the ideal
    $\mathfrak{a} \subseteq \Bbbk[X,U]$. We represent $\mathfrak{a}' = \mathfrak{a} \cap \Bbbk[u_{1},\ldots,u_{m}]$.
    \Ensure A comprehensive border basis $B$ of $\mathfrak{a}$
    \State $\mathcal{B}_{1} = \{(\gamma, \mathcal{O}_{i}, B_{i})\}$ such that $\gamma$ is a condition belonging to
    $\bar{\Bbbk}^m - \mathcal{V}(\mathfrak{a}')$.
    \State $\mathcal{B}' = \mathcal{B} \setminus \mathcal{B}_{1}$
    \State Let $f_{\mathfrak{a}'} = \prod\limits_{\gamma \in \mathcal{V}(\mathfrak{a}')} f_{\gamma}$
    \State\label{MarkedPoly} Replace each $B_{i}$ belonging to $(\gamma_{i}, \mathcal{O}_{i}, B_{i}) \in \mathcal{B}'$ by $B_{i_{mark}}$, where $B_{i_{mark}}$
    is marked version of $B_{i}$ such that, border form of each polynomial in $B_{i}$ is marked with respect to $\mathcal{O}_{i}$.
    \State\label{ConditionalPoly} Replace each $B_{i_{mark}}$ belonging to $(\gamma_{i}, \mathcal{O}_{i}, B_{i_{mark}}) \in \mathcal{B}'$ by $B_{i_{CBB}}$
    \[ B_{i_{CBB}} = \{ (f_{\mathfrak{a}} / f_{\gamma_{i}}) (f) | f \in B_{i_{mark}} \} \]
    \State $B = \bigcup\limits_{B_{i_{CBB}} \in \mathcal{B}'} B_{i_{CBB}} \cup \{B_{i} | B_{i} \in \mathcal{B}_{1}\}$
    \State Return B (CBB)
 \end{algorithmic}
\end{algorithm}

Correctness of the algorithm can be seen from the nature of the
vanishing polynomial under a specialization. 
For a specialization $\gamma_{i} \in \mathcal{V}(\mathfrak{a}')$ such that, $(\gamma_{i},
\mathcal{O}_{i}, B_{i}) \in \mathcal{B}'$ the vanishing polynomial 
$f_{\gamma_{i}}$ is multiplied to all the conditional border bases in
$\mathcal{B}'$ except for $B_{i}$ (Step~\ref{ConditionalPoly}). This implies for all  
$B_{j} \in \mathcal{B}', j \neq i$, $B_{j_{CBB}}$ vanishes under $\gamma_{i}$ and also
$B_{\ell}$ corresponding to $\mathcal{B}_{1}$ vanishes under
$\gamma_{i}$. 
For $\gamma_{\ell} \notin \mathcal{V}(\mathfrak{a}')$ such that
$(\gamma_{\ell},\mathcal{O}_{\ell}, B_{\ell}) \in \mathcal{B}_{1}$
we have $\langle \sigma(B_{\ell}) \rangle =
\langle1\rangle$ for all $\sigma \in \gamma_{\ell}$.
 
One must note here that the border system and hence the CBB
constructed in Algorithm~\ref{Border System Algorithm} and  
Algorithm~\ref{Comprehensive border bases Algorithm} are not
faithful. As the CBB for an ideal is computed 
from the border system, if the border system is faithful, the
CBB constructed also be faithful. We can obtain the faithful border
system by restricting 
polynomial operations (i.e. avoiding division or multiplication by inverses)
in $\Bbbk[U][X]$ in the colored border bases algorithm
(Algorithm~\ref{Border System Algorithm} Step~\ref{coloredBBStep}).
This is similar to the technique followed in  
\citep{Weispfenning:1992:ComprehensiveGrobnerbases}, where a `modified'
definition of reduction and $S$-polynomial is used to avoid
divisions. 

For construction of border bases under a specialization from the
CBB computed in Algorithm~\ref{Comprehensive
  border bases Algorithm}, we have to calculate the order ideal
$\mathcal{O}$ as $\mathbb{T}^{n} \setminus \langle M \rangle$, where M
is the set of all marked monomials (Algorithm~\ref{Comprehensive
  border bases Algorithm}, Step~\ref{MarkedPoly}) that do not vanish on specialization.
  The correctness of $\mathcal{O}$ is implied by the construction of CBB shown above.

A note on identifying whether a given CBB correspond to any term
order or not. If a border basis correspond to a term order if and only
if it contains a reduced Gr\"{o}bner bases as the subset. These border
bases are hence also the Gr\"{o}bner bases for the corresponding term
order.


Let $B \in \Bbbk[U][X]$ be a CBB and $\sigma \in \bar\Bbbk^m$ be a
specialization. One can verify whether $\sigma(B) \subseteq
\bar\Bbbk[X]$ is a Gr\"{o}bner bases or not using 
algorithm given in \citep{sturmfels:1996:grobner}. If it is a Gr\"{o}bner basis
then it implies that $B$ correspond to a term order else $B$ is a CBB
that do not correspond to any term order. We must note that the nature
of CBB generated depends on the border bases algorithm (colored
version) used in Algorithm~\ref{Border System Algorithm}. If we
already know the nature of border bases algorithm used then CBB
generated will also be of the same nature.  
\section{Relation between CGB and CBB}
\label{Realtion between CGB and CBB}
We establish the following result that is crucial for studying the
relation between CGB and CBB. 
\begin{proposition}
\label{Border Form ideal and Leading term ideal realtion}
For a given term order and an ideal, border form ideal is same as
leading term ideal. 
\end{proposition}
\proof
For an ideal $\mathfrak{a} \subseteq \Bbbk[X]$ and a given term order
$\leq$, assume $G \subset \mathfrak{a}$ be a reduced Gr\"{o}bner
basis. 
Let $\mathcal{O}$ be the $\Bbbk$-vector space basis of
$\Bbbk[X]/\mathfrak{a}$ corresponding to $G$. Then there exists a
unique 
$\mathcal{O}$-border basis $B$ for $\mathfrak{a}$ corresponding to
$\leq$. Now from Gr\"{o}bner basis $G$, $\mathcal{O}$ is given by 
${\mathbb{T}}^n \setminus \langle \operatorname{LT}(G) \rangle = {\mathbb{T}}^n \setminus \langle \operatorname{LT}_{\leq}(\mathfrak{a}) \rangle$.
And from $\mathcal{O}$-border bases $B$, $\mathcal{O}$ is given by ${\mathbb{T}}^n \setminus \langle BF_{\mathcal{O}}(\mathfrak{a}) \rangle$.
Hence, for a given term order $\leq$ we have a unique $\Bbbk$-vector space basis $\mathcal{O}$ such that
$\operatorname{LT}_{\leq}(\mathfrak{a}) = BF_{\mathcal{O}}(\mathfrak{a})$.
\endproof

With the above proposition, one must note that the existential proof
of CBB is intuitive from CGB only in the case, where the border bases 
correspond to the Gr\"{o}bner bases (term order) and not otherwise.



$\bar{\Bbbk}$. 

Let $\mathfrak{a} \subseteq \Bbbk[X,U]$ be an ideal and a term ordering 
$\leq$ with an elimination order satisfying $X \geq U$. 
$\operatorname{LM}_{\gamma}(g)$ denotes conditional leading red term of the
polynomial $g$ under the condition $\gamma$, and $\operatorname{LM}_{\gamma}(G)$ 
is the set of conditional leading red terms of the polynomials in $G$
under $\gamma$. From the property of Gr\"{o}bner system 
and hence reduced Gr\"{o}bner system, for each tuple
$\{(G,\gamma)\} \in \mathcal{G}$, conditional leading red terms of the
polynomials in $G$ under the condition $\gamma$ is uniquely
determined. 
The process of computation of CBB from CGB is described below. 
\vspace{5mm}
CGB
$\xrightarrow[]{Weispfenning (1992)}$
Gr\"{o}bner system,
$\mathcal{G} \xrightarrow[]{Weispfenning (1992)}$ 
Reduced Gr\"{o}bner system
$\xrightarrow[]{Algorithm~\ref{reduced GS to BS}}$
border system,
$\mathcal{B} \xrightarrow[]{Algorithm~\ref{Comprehensive border bases Algorithm}} CBB$ 
\vspace{5mm}

\begin{algorithm}[ht]
    \caption{Reduced Gr\"{o}bner system to border system}\label{reduced GS to BS}
    \begin{algorithmic}[1]
    \Require Reduced Gr\"{o}bner system $\mathcal{G} = \{(G,\gamma)\}$ w.r.t $\leq$ term order.
    \Ensure Border system $\mathcal{B}$
	\State $\mathcal{B} = \emptyset$
	\While{$\mathcal{G} \neq \emptyset$}
	    \State Select any $(G_{i},\gamma_{i}) \in \mathcal{G}$
	    \State $\mathcal{G} = \mathcal{G} \setminus \{(G_{i},\gamma_{i})\}$
	    \State\label{startConvert} Compute $\mathcal{O}_{i} = \mathbb{T}^n \setminus \langle\{ LM_{\gamma_{i}}(G_{i})\}\rangle$
	    \State Compute $\partial\mathcal{O}_{i}$
	    \State $\partial\mathcal{O}'_{i} = \partial\mathcal{O}_{i} \setminus \{\operatorname{LM}_{\gamma_{i}}(G_{i})\}$
	    \While {$\partial\mathcal{O}'_{i}$ is not empty}
		    \State\label{minOrderTerm} Select a monomial $x^{\alpha} \in \partial\mathcal{O}'_{i}$
		    \State $\partial\mathcal{O}'_{i} = \partial\mathcal{O}'_{i} \setminus x^{\alpha}$
		    \State Get the variable $x_{\ell}$, such that $x^{\alpha} = x_{\ell}LM_{\gamma_{i}}(g_{j})$ for some $g_{j} \in G_{i}$.
		    \State\label{borderTermPoly} Consider $b = x_{\ell}g_{j} = cx^{\alpha} + p$ \Comment $c \in \Bbbk$
		    \State Reduce $p$ to $p'$, such that $p'$ is irreducible with $G_{i}$ relative to $\gamma_{i}$
		    \State $b = x^{\alpha} + p'$
		    \State $B_{i} = B_{i} \cup b$
	    \EndWhile\label{endConvert}
	    \State $\mathcal{B} = \mathcal{B} \cup \{(\mathcal{V}(\gamma_{i}), \mathcal{O}_{i}, B_{i})\}$
	\EndWhile
	\State Return $\mathcal{B}$
    \end{algorithmic}
\end{algorithm}

As for each tuple $\{(G,\gamma)\} \in \mathcal{G}$, for any
specialization $\sigma$ corresponding to $\gamma$, $\sigma(G)$ is a
zero dimensional ideal and so we have a finite order ideal for $\langle \sigma(G)\rangle$ corresponding to each tuple. The finiteness of the
order ideal shows the termination of the inner while loop. The reduced Gr\"{o}bner system calculated from Gr\"{o}bner system is a finite set of
tuples, which shows the termination of the outer while loop. Hence, Algorithm~\ref{reduced GS to BS} terminates.

It is easy to see that for each tuple $(G,\gamma) \in \mathcal{G}$
(reduced Gr\"{o}bner system) the conditional leading term ideal is
same for all specializations $\sigma$ corresponding to the
condition $\gamma$, and so the border form ideal will remain same for
the ideal  formed by $\mathcal{G}$ under any such specialization. Due
to this fact the conditional reduced Gr\"{o}bner bases $G$ under
$\gamma$ maps to conditional scalar border bases $B$ under
$\gamma$ (Proposition~\ref{Border Form ideal and Leading term ideal realtion}).
So an element in $\mathcal{G}$ maps to an element in
$\mathcal{B}$. Each tuple $(G,\gamma) \in \mathcal{G}$ under a
specialization $\sigma$ corresponding to the condition $\gamma$ forms
a scalar reduced Gr\"{o}bner bases (reduced Gr\"{o}bner bases except for the
monic leading coefficient). We have to convert each of this
conditional scalar reduced Gr\"{o}bner bases to conditional scalar
border bases. Step~\ref{startConvert} to Step~\ref{endConvert} is the
colored version of the standard Gr\"{o}bner bases to border bases
algorithm. It therefore computes conditional scalar border bases from scalar
reduced Gr\"{o}bner bases. Rest of the algorithm uses
Step~\ref{startConvert} to Step~\ref{endConvert} to map each
element in $\mathcal{G}$ to an element in $\mathcal{B}$.  

One must note that comprehensive border bases calculated w.r.t to some
term order, will also be the CGB. This is obvious by the fact that for
a given term order, border bases is a superset of reduced Gr\"{o}bner
bases.  

\section{Comprehensive Border bases and regular rings}\label{CBB over regular rings}
Regular rings (commutative von Neumann regular rings) can be viewed as a
certain subdirect products of the fields. The close relation between
comprehensive Gr\"{o}bner bases over fields and non-parametric
Gr\"{o}bner bases over von Neumann regular rings was shown in
\citep{weispfenning:2002:regularRings,
  weispfenning:2006:regularRings}. Any algorithm for Gr\"{o}bner bases
over fields can be modified to compute Gr\"{o}bner bases over von
Neumann regular rings, these Gr\"{o}bner bases over von Neumann
regular rings also give us comprehensive Gr\"{o}bner
bases.

Computation of the comprehensive border bases can also be done
through computing border bases over von Neumann regular rings. All the
operations and the ideas come exactly from the computation of Gr\"{o}bner
bases over regular rings. We must make sure that the final output of
border bases over regular rings have all the boolean closed
polynomials. The difference comes in the structure of the order ideal
over regular rings. We know that the order ideal contains all the monic
terms (monomials), but the definition of monicness
changes over regular rings. 

Let $R$ be a regular ring and $a \in R$, then the element $a^*$, such that  $a.a^* = a$, is called the idempotent element of a.
Let $f \in R[X]$ be a polynomial over regular ring, $R$. 
\begin{definition}[\cite{sato:1998:regularRings}]
A polynomial $f$ is called monic if it satisfies $\operatorname{LC}(f)$ = $(\operatorname{LC}(f))^*$ 
\end{definition}
So the order ideal will have monic coefficients coming from $R$. This can be seen as the coordinate wise order ideal for
border bases over regular rings (coordinate wise, order ideal and the border bases over the field). 

Also, the polynomial coefficients from the polynomial ring $\Bbbk[U]$ can be extended to computable ring of terraces
\citep{suzuki:2003:terracesCGB} so that we get the closure under addition, multiplication and inverses. Ring of terraces
becomes von Neumann regular rings and the terrace arithmetic is well studied in \citep{suzuki:2003:terracesCGB}. Output
of the Gr\"{o}bner bases computed with terraces as the coefficients is similar to the output of the Gr\"{o}bner system, in the sense
that, we know the corresponding Gr\"{o}bner bases for each possible substitution of the parameters. So for border bases we can
associate an order ideal with each possible border bases output.

\section{Example}
\label{Example section}
For the colored version of a border bases algorithm, we modify the border basis algorithm given in
\citep{KehreinKreuzer:2006:ComputingBB} and use the deglex term ordering.
Consider a simple example for the algorithms given above.

{\bf Example:}
F= $\{x^2-z^2-6x+4z+5,\ 3y^2+z^2-12y-4z+12,\ z^3-8z^2+19z-12,\ x^2z^3-8x^2z^2+19x^2z+xz^2-12x^2-4xz-z^2+3x+4z-3,\
x^2z^3-8x^2z^2+19x^2z+yz^2-12x^2-4yz-2z^2+3y+8z-6 \}$

First, we compute a border system $\mathcal{B}$, using Algorithm~\ref{Border System Algorithm}.
\begin{enumerate}
 \item $\mathcal{V}(\mathfrak{a}') = \{4, 3, 1\}$, where $\mathfrak{a}' = \langle z^3-8z^2+19z-12\rangle$
 \item $F' = \{x^2-6x-(z^2-4z-5)$,\newline
		  $3y^2-12y+(z^2-4z+12)$,\newline
		  $(z^3-8z^2+19z-12)x^2+(z^2-4z+3)x-(z^2-4z+3)$,\newline
		  $(z^3-8z^2+19z-12)x^2+(z^2-4z+3)y-2(z^2-4z+3)\}$
 \item[(3)] {\bf pass 1 for specialization $z = 4$}
	\begin{enumerate}
	\item[(4)]$F'_{\sigma} = \{x^2_{R}-6x_{R}-(z^2-4z-5)_{R}$,\newline
		  $3y^2_{R}-12y_{R}+(z^2-4z+12)_{R}$,\newline
		  $(z^3-8z^2+19z-12)x^2_{G}+(z^2-4z+3)x_{R}-(z^2-4z+3)_{R}$,\newline
		  $(z^3-8z^2+19z-12)x^2_{G}+(z^2-4z+3)y_{R}-2(z^2-4z+3)_{R}\}$
	\item[(5)]$\mathcal{O} = \{1\}$,\newline
			      $\partial\mathcal{O} = \{x, y\}$,\newline
			      $\operatorname{\mathcal{O}-BB}_{z=4}: \{(z-4)x^2_{G}+x_{R}-1_{R},\ (z-4)x^2_{G}+y_{R}-2_{R}\}$
	\item[(6)]$\mathcal{O} = \{1\}$,\newline
			      $\partial\mathcal{O} = \{x,y\}$,\newline
			      $\operatorname{\mathcal{O}-BB}_{z=4}: \{(z-4)x^2+x-1,\ (z-4)x^2+y-2\}$
	\item[(7)] $\mathcal{B} = \{(\{4\},\{x,y\},\{(z-4)x^2+x-1,\ (z-4)x^2+y-2\})\}$
	\end{enumerate}
	
 \item[(3)]{\bf pass 2 for specialization $z = 3$}
	\begin{enumerate}
	 \item[(4)]$F'_{\sigma} = \{x^2_{R}-6x_{R}-(z^2-4z-5)_{R}$,\newline
		  $3y^2_{R}-12y_{R}+(z^2-4z+12)_{R}\}$
	 \item[(5)]$\mathcal{O} = \{1, x, y, xy\}$,\newline
		    $\partial\mathcal{O} = \{x^2,y^2, x^2y, xy^2\}$,\newline
		    $\operatorname{\mathcal{O}-BB}_{z=3}: \{x^2_{R}-6x_{R}-(z^2-4z-5)_{R},\ y^2_{R}-4y_{R}+1/3(z^2-4z+12)_{R},\ x^2y_{R}-6xy_{R}-y(z^2-4z-5)_{R},\ xy^2_{R}-4xy_{R}+1/3x(z^2-4z+12)_{R}\}$
	 \item[(6)]$\mathcal{O} = \{1, x, y, xy\}$,\newline
		    $\partial\mathcal{O} = \{x^2,y^2, x^2y, xy^2\}$,\newline
		    $\operatorname{\mathcal{O}-BB}_{z=3}: \{x^2-6x-(z^2-4z-5),\ y^2-4y+1/3(z^2-4z+12),\ x^2y-6xy-y(z^2-4z-5),\ xy^2-4xy+1/3x(z^2-4z+12)\}$
	 \item[(7)]$\mathcal{B} = \{(\{4\},\{1\},\{(z-4)x^2+x-1,\ (z-4)x^2+y-2\})$,
			  $(\{3\},\{1, x, y, xy\}, \{x^2-6x-(z^2-4z-5),\ y^2-4y+1/3(z^2-4z+12),\ x^2y-6xy-y(z^2-4z-5),\ xy^2-4xy+1/3x(z^2-4z+12)\})\}$
	\end{enumerate}
	
 \item[(3)]{\bf pass 3 for specialization $z = 1$}
	 \begin{enumerate}
	  \item[(4)]$F'_{\sigma} = \{x^2_{R}-6x_{R}-(z^2-4z-5)_{R}$,\newline
		  $3y^2_{R}-12y_{R}+(z^2-4z+12)_{R}\}$
	 \item[(5)]$\mathcal{O} = \{1, x, y, xy\}$,\newline
		    $\partial\mathcal{O} = \{x^2,y^2, x^2y, xy^2\}$,\newline
		    $\operatorname{\mathcal{O}-BB}_{z=3}: \{x^2_{R}-6x_{R}-(z^2-4z-5)_{R},\ y^2_{R}-4y_{R}+1/3(z^2-4z+12)_{R},\ x^2y_{R}-6xy_{R}-y(z^2-4z-5)_{R},\ xy^2_{R}-4xy_{R}+1/3x(z^2-4z+12)_{R}\}$
	 \item[(6)]$\mathcal{O} = \{1, x, y, xy\}$,\newline
		    $\partial\mathcal{O} = \{x^2,y^2, x^2y, xy^2\}$,\newline
		    $\operatorname{\mathcal{O}-BB}_{z=3}: \{x^2-6x-(z^2-4z-5),\ y^2-4y+1/3(z^2-4z+12),\ x^2y-6xy-y(z^2-4z-5),\ xy^2-4xy+1/3x(z^2-4z+12)\}$	 
	 \item[(7)]$\mathcal{B} = \{(\{4\},\{1\},\{(z-4)x^2+x-1,\ (z-4)x^2+y-2\})$,
			  $(\{3\},\{1, x, y, xy\}, \{x^2-6x-(z^2-4z-5),\ y^2-4y+1/3(z^2-4z+12),\ x^2y-6xy-y(z^2-4z-5),\ xy^2-4xy+1/3x(z^2-4z+12)\})$,
			  $(\{1\},\{1, x, y, xy\}, \{x^2-6x-(z^2-4z-5),\ y^2-4y+1/3(z^2-4z+12),\ x^2y-6xy-y(z^2-4z-5),\ xy^2-4xy+1/3x(z^2-4z+12)\})\}$
	 \end{enumerate}
 \item[(8)] end for loop
 \item[(9)] $\mathcal{B} = \{(\{4\},\{1\},\{(z-4)x^2+x-1,\ (z-4)x^2+y-2\})$,\newline
			  $(\{3\},\{1, x, y, xy\}, \{x^2-6x-(z^2-4z-5),\ y^2-4y+1/3(z^2-4z+12),\ x^2y-6xy-y(z^2-4z-5),\ xy^2-4xy+1/3x(z^2-4z+12)\})$,\newline
			  $(\{1\},\{1, x, y, xy\}, \{x^2-6x-(z^2-4z-5),\ y^2-4y+1/3(z^2-4z+12),\ x^2y-6xy-y(z^2-4z-5),\ xy^2-4xy+1/3x(z^2-4z+12)\})$,\newline
			  $(\{\bar{\Bbbk} \setminus \{4, 3, 1\}\},\emptyset,\{z^3-8z^2+19z-12\})\}$
  \item[(10)] $\mathcal{B} = \{(\{4\},\{1\},\{(z-4)x^2+x-1,\ (z-4)x^2+y-2\})$,\newline
			  $(\{3, 1\},\{1, x, y, xy\}, \{x^2-6x-(z^2-4z-5),\ y^2-4y+1/3(z^2-4z+12),\ x^2y-6xy-y(z^2-4z-5),\ xy^2-4xy+1/3x(z^2-4z+12)\})$,\newline			  
			  $(\{\bar{\Bbbk} \setminus \{4, 3, 1\}\},\emptyset,\{z^3-8z^2+19z-12\})\}$
\end{enumerate}

We should note that the colored version of any border bases algorithm is just used as a plugin for Algorithm~\ref{Border System Algorithm}
in step~\ref{coloredBBStep}. As we have used the border bases algorithm which uses a degree compatible term ordering, we can also
verify the output above with the CGB output.
We can compare our output here because of Theorem~\ref{Unique Border and reduced Grobner bases} and
Section~\ref{Realtion between CGB and CBB}. 

\textbf{Note:} The border system output depends on the colored version of the border bases algorithm we use. If we use the border bases
algorithm which does not correspond to any Gr\"{o}bner bases then we
can generate comprehensive border bases for which we can't verify the 
output with CGB output.\newline

Let us see the output of comprehensive border bases algorithm. We will show the marked term of a polynomial by subscript m.
\begin{enumerate}
 \item $\mathcal{B}_{1} = \{(\{\bar{\Bbbk} \setminus \{4, 3, 1\}\},\emptyset,\{z^3-8z^2+19z-12\})\}$
 \item $\mathcal{B}' = \{(\{4\},\{1\},\{(z-4)x^2+x-1,\ (z-4)x^2+y-2\})$,\newline
			  $(\{3, 1\},\{1, x, y, xy\}, \{x^2-6x-(z^2-4z-5),\ y^2-4y+1/3(z^2-4z+12),\ x^2y-6xy-y(z^2-4z-5),\ xy^2-4xy+1/3x(z^2-4z+12)\})\}$
 \item Let $f_{\mathfrak{a}} = (z-1)(z-3)(z-4)$
 \item $\mathcal{B}' = \{(\{4\},\{1\},\{(z-4)x^2+x_{m}-1,\ (z-4)x^2+y_{m}-2\})$,\newline
			  $(\{3, 1\},\{1, x, y, xy\}, \{x^2_{m}-6x-(z^2-4z-5),\ y^2_{m}-4y+1/3(z^2-4z+12),\ x^2y_{m}-6xy-y(z^2-4z-5),\ xy^2_{m}-4xy+1/3x(z^2-4z+12)\})\}$
 \item $\mathcal{B}' = \{(\{4\},\{1\},\{(z-1)(z-3)((z-4)x^2+x_{m}-1),\ (z-1)(z-3)((z-4)x^2+y_{m}-2)\})$,\newline
			  $(\{3, 1\},\{1, x, y, xy\}, \{(z-1)(z-3)(x^2_{m}-6x-(z^2-4z-5)),\ (z-1)(z-3)(y^2_{m}-4y+1/3(z^2-4z+12)),\ (z-1)(z-3)(x^2y_{m}-6xy-y(z^2-4z-5)),\ (z-1)(z-3)(xy^2_{m}-4xy+1/3x(z^2-4z+12))\})\}$
 \item $B = \{(z-1)(z-3)((z-4)x^2+x_{m}-1)$,\newline $(z-1)(z-3)((z-4)x^2+y_{m}-2)$,\newline $(z-4)(x^2_{m}-6x-(z^2-4z-5))$,\newline $(z-4)(y^2_{m}-4y+1/3(z^2-4z+12))$,\newline $(z-4)(x^2y_{m}-6xy-y(z^2-4z-5))$,\newline
		  $(z-4)(xy^2_{m}-4xy+1/3x(z^2-4z+12))$,\newline $z^3-8z^2+19z-12\}$
\end{enumerate}


Lets calculate $\operatorname{\mathcal{O}-BB}$ for the specialization $z = 4$ and $z = 7$.
\begin{enumerate}
 \item $z = 4\ \mathrm{i.e.}\ \sigma_{4}$\newline
 $\sigma_{4}(B) = \{x_{m}-1,\ y_{m}-2\} = \{x-1,\ y-1\}$\newline
 $\mathcal{O} = \mathbb{T}^n \setminus \langle x, y \rangle = \{1\}$\newline
 We can verify that the $\sigma_{4}(B)$ is $\operatorname{\mathcal{O}-BB}$ of $\sigma_{4}(\mathfrak{a})$.
 \item $z = 7\ \mathrm{i.e.}\ \sigma_{7}$\newline
 $\sigma_{7}(B) = \{3x^2+x_{m}-1$,\newline $3x^2+y_{m}-2$,\newline $x^2_{m}-6x-(z^2-4z-5)$,\newline $y^2_{m}-4y+1/3(z^2-4z+12)$,\newline $x^2y_{m}-6xy-y(z^2-4z-5)$,\newline
		  $xy^2_{m}-4xy+1/3x(z^2-4z+12), 1 \}$\newline
		  $= \{1\}$\newline
 $\mathcal{O} = \mathbb{T}^n \setminus \langle 1 \rangle = \emptyset$\newline
 We can verify that the $\sigma_{7}(B)$ is $\operatorname{\mathcal{O}-BB}$ of $\sigma_{7}(\mathfrak{a})$.
\end{enumerate}

\section{Concluding remarks}
\label{Concluding remarks}
The theory of comprehensive Gr\"obner bases  is an important tool
for the studying parametric ideals. 
Applications of comprehensive Gr\"obner bases include ideal membership depending upon 
parameters, elimination of quantifier-blocks in algebraically closed
fields, deformation of residue algebras, geometric theorem proving and 
many more.

In this paper we proposed the notion of comprehensive border bases  for
zero dimensional ideals. We established the existence of comprehensive
border bases that need not correspond to any term order and hence to
any comprehensive Gr\"obner bases. We also propose an algorithm to
compute  comprehensive border bases and study its relation with
comprehensive Gr\"obner bases. 
The main aim of the proposed comprehensive border bases is to bring
the features of border bases computation in the studies of
zero-dimensional parametric ideals. 




\section*{Acknowledge}
The authors would like to acknowledge Sushma Palimar, Maria Francis and Nithish Pai for the useful discussions.

{\footnotesize
 \bibliographystyle{jtbnew}
 \bibliography{refs}
}

\end{document}